\documentclass[conference]{IEEEtran}
\IEEEoverridecommandlockouts
\usepackage{graphicx}
\usepackage[caption=false,textfont=sf]{subfig}
\usepackage{changes}
\usepackage{amsmath}
\usepackage{graphicx}
\usepackage{multicol}
\usepackage{amssymb}
\usepackage{subfig}
\usepackage{lineno}
\usepackage{makecell}
\usepackage{indentfirst}
\usepackage{bm}
\usepackage{color}
\usepackage{cite}
\usepackage{epstopdf}
\usepackage[noend]{algpseudocode}
\usepackage{algorithmicx,algorithm}
\usepackage{amssymb}
\usepackage{xcolor}
\usepackage{tikz}
\usepackage{mathtools}

\newtheorem{mypro}{{Proposition}}
\begin{document}
	
	\title{Intelligent Reflecting Surface-Aided Maneuvering Target Sensing: True Velocity Estimation}
	\IEEEspecialpapernotice{(Invited Paper)}
	\author{Lei~Xie,~Xianghao Yu,~and Shenghui Song\\
		{Dept. of ECE, The Hong Kong University of Science and Technology, Hong Kong}\\
		{Email: $\{$eelxie, eexyu, eeshsong$\}$@ust.hk}
	\thanks{This work was supported by the Shenzhen Science and Technology Innovation Committee under Grant SGDX20210823103201006 and the HKUST-BDR Joint Research Institute under Grant OKT22EG04.}}

	\maketitle
	
	\begin{abstract}
		Maneuvering target sensing will be an important service of future vehicular networks, where precise velocity estimation is one of the core tasks.
		To this end, the recently proposed integrated sensing and communications (ISAC) provides a promising platform for achieving accurate velocity estimation.
		However, with one mono-static ISAC base station (BS), only the radial projection of the true velocity can be estimated, which causes serious estimation error.
		In this paper, we investigate the estimation of the true velocity of a maneuvering target with the assistance of an intelligent reflecting surface (IRS).
		We propose an efficient velocity estimation algorithm by exploiting the two perspectives from the BS and IRS to the target. We propose a two-stage scheme where the true velocity can be recovered based on the Doppler frequency of the BS-target link and BS-IRS-target link. Experimental results validate that the true velocity can be precisely recovered and demonstrate the advantage of adding the IRS.
	\end{abstract}
	
	\begin{IEEEkeywords}
		Intelligent reflecting surface, integrated sensing and communications, Doppler effect, true velocity estimation.
	\end{IEEEkeywords}

	\section{Introduction}

	Vehicular to everything (V2X) communications is becoming a key enabling technology for many thrilling applications, e.g., smart city and intelligent transportation systems.
	However, V2X poses stringent requirements on both sensing accuracy and communication quality.
	To this end, the recently proposed integrated sensing and communication (ISAC) provides a promising platform that amalgamates the abilities of sensing and communication, especially for 5G and beyond systems  \cite{liu2020joint,xie2022perceptive,xie2022collaborative}.
	However, high-accuracy beam tracking is necessary to compensate for the path loss imposed by the high-frequency signals and guarantee the quality-of-service (QoS) for V2X, particularly in high-mobility vehicular scenarios \cite{liu2020radar}.

	To achieve accurate beam-tracking, precise velocity estimation is required.  
	Doppler effect, which refers to the change in frequency of an electromagnetic wave caused by the relative motion between the observer and the wave source, is typically utilized for velocity estimation. 
	Existing works on velocity estimation mainly adopted matched filters (MFs) to estimate the motion parameters \cite{6847156,lim2015two,liu2020radar}. 
	Micro-Doppler frequency estimation was considered in \cite{6847156} for orthogonal frequency division multiplexing (OFDM) radar systems. In \cite{lim2015two}, the authors proposed to mitigate the inter-carrier interference caused by the Doppler effect to improve the performance of Doppler estimation by oversampling.
	A predictive beamforming design was investigated for an ISAC system in \cite{liu2020radar}. In general, the performance of the MF-based method is limited by the inherent grid issue. 
	To address this problem, a joint range-velocity estimation was proposed in \cite{dokhanchi2019mmwave} based on the multiple signal classification (MUSIC) algorithm in OFDM radar systems, which achieves a higher resolution than the MF-based methods.
	However, due to the nature of the Doppler effect, the conventional mono-static ISAC BS can only measure the radial projection of the true velocity, and thus the true velocity cannot be uniquely recovered. One solution is to obtain another perspective toward the maneuvering target so that the true velocity can be uniquely determined.
	
	In recent years, intelligent reflecting surfaces (IRSs) have been introduced to enhance the performance of both communication and sensing \cite{yu2021irs,9690635,he2020adaptive}. 
	Some research efforts have been made to IRS-aided systems to tackle user equipment (UE) mobility \cite{matthiesen2020intelligent,huang2021transforming,keykhosravi2022ris}. The authors in \cite{matthiesen2020intelligent} considered a continuous model for IRS-aided satellite communication, based on which the IRS phase shifters are optimized to simultaneously maximize the received
	power and minimize the delay and Doppler spread. 
	In \cite{huang2021transforming}, a two-stage protocol for channel estimation was proposed for an IRS-aided high-mobility communication system, which deploys an IRS at a high-speed vehicle to enhance the QoS for passengers by mitigating the Doppler effect.
	In \cite{keykhosravi2022ris}, the authors considered the localization problem in an IRS-aided single-input single-output (SISO) system by taking into account the UE mobility and spatial wideband effects.
	However, the potential of IRSs for true velocity estimation has not been fully understood.

	In this paper, by leveraging the additional perspective  provided by the IRS, we study the true velocity estimation in an IRS-aided ISAC system. 
	By exploiting the geometric information, the true velocity can be recovered based on the Doppler frequency of the BS-target link and BS-IRS-target link\footnote{For simplicity, we will refer to the BS-target link and BS-IRS-target link as the direct link and IRS link, respectively.}. 
	To estimate the Doppler frequency of both links, we propose a two-stage scheme. In the first stage, a coarse estimation of radial projection of the true velocity is obtained by turning off the IRS. In the second stage, a gridless estimation for the Doppler frequency of the direct link and IRS link is performed by exploiting the method of direction estimation (MODE) \cite{57542}. 
	Compared with existing gridless estimation methods such as root-multiple signal classification (root-MUSIC) \cite{1143830} and estimation of signal parameters via rotation invariance techniques (ESPRIT) \cite{32276}, the MODE approach is shown to provide more accurate parameter estimation \cite{57542}.

	\section{Preliminary}
\begin{figure}[!t]
		\centering
		\includegraphics[width=2.8in]{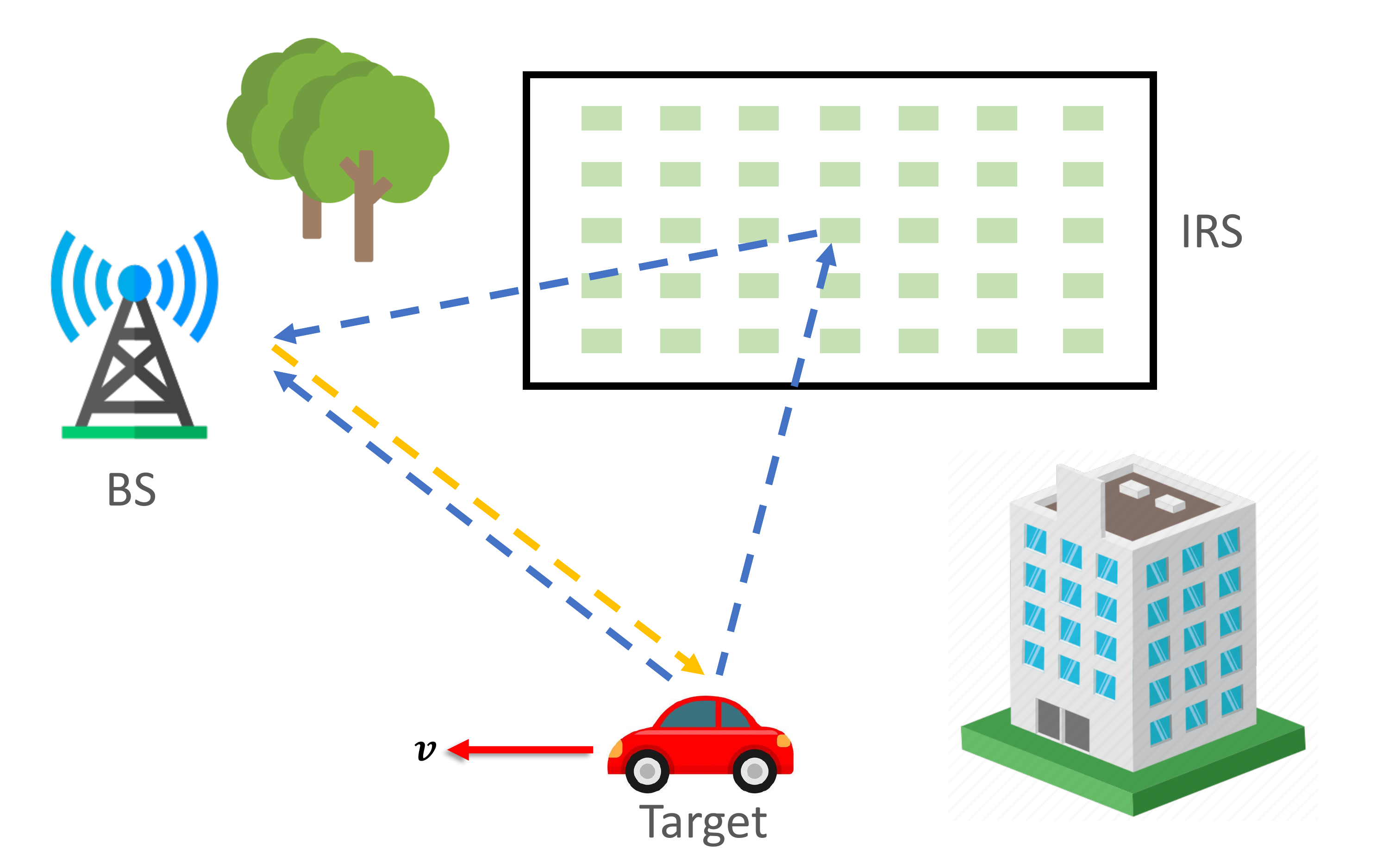}
		\caption{Illustration of the considered IRS-aided ISAC system. }
		\label{fig_sys}
	\end{figure}

	Consider an ISAC system, as illustrated in Fig. \ref{fig_sys}, where the velocity vector of the point-like maneuvering target, denoted by $\mathbf{v}$\footnote{The velocity vector is defined by $\mathbf{v}=\left[|\mathbf{v}|\cos \theta_v,|\mathbf{v}|\sin \theta_v\right]^\mathrm{T}$, where $\theta_v$ represents the direction of velocity as illustrated in Fig. \ref{fig_illuDop}.}, is estimated by one BS with the help of an IRS. The BS and IRS are equipped with a uniform linear array (ULA) with $N$ and $M$ antennas, respectively.

	\subsection{Signal Model}

	Assume that the directions of the target with respect to the  BS and IRS, denoted by  ${\theta}_{tb}$ and ${\theta}_{it}$ in Fig. \ref{fig_illuDop}(b), are well estimated in advance.
	To estimate the true velocity, the BS transmits a probing beam
	\begin{equation}
		\begin{split}
			\mathbf{f}&=\mathbf{a}\left(\theta_{tb}\right)\triangleq
			\left[1,e^{j\frac{2\pi d}{\lambda} \cos \theta_{tb}},\cdots, e^{j\frac{2\pi d(N-1)}{\lambda} \cos \theta_{tb}}\right]^{\mathrm{T}},
		\end{split}
	\end{equation}
	where $d$ and $\lambda$ represent the inner spacing of the ULA and the wavelength, respectively. Here, $\mathbf{a}(\cdot)$ represents the steering vector of the antenna array of the BS towards the target.  Note that due to the use of the narrow beam, the effect of the non-light-of-sight (NLoS) path is considered negligible.

	The echo of the $k$-th pilot symbol $s_k$ received by the BS is given by 
	\begin{equation}\label{yldef}
		\begin{split}
			&\mathbf{y}_{r,k}= \underbrace{\alpha_d e^{j2\pi \mu_{d} (k-1) T_{s}} \mathbf{a}(\theta_{tb}) \mathbf{a}^{\mathrm{H}}(\theta_{tb}) \mathbf{f}s_k}_{\text{direct link}}\\
			&+ \underbrace{ \alpha_r e^{j2\pi \mu_{r} (k-1) T_{s}} \mathbf{G} \mathbf{\Psi} \mathbf{b}(\theta_{it}) \mathbf{a}^{\mathrm{H}}(\theta_{tb}) \mathbf{f}s_k}_{\text{IRS link}}+\mathbf{p}_{r,k},
		\end{split}
	\end{equation}
	where 	
	\begin{equation}
		\begin{split}
			\mathbf{b}\left({\theta}_{it}\right)\triangleq
			\left[1,e^{j\frac{2\pi d}{\lambda} \cos {\theta}_{it}},\cdots, e^{j\frac{2\pi d (M-1)}{\lambda} \cos {\theta}_{it}}\right]^{\mathrm{T}}
		\end{split}
	\end{equation}
	represents the response vector of the IRS, 
	$\mathbf{G}$ denotes the channel between BS and IRS, which is modeled as a Rician channel comprising a light-of-sight (LoS) path and a number of NLoS paths, 
	$\mathbf{\Psi}$
	denotes the phase shifter which is designed to align toward the target\footnote{The optimized design of phase shifter will be left as future work.}, $\alpha_d$ and $\alpha_r$ 
	denote the complex channel gain, accounting	for the path loss and target reflectivity\footnote{The fluctuations of target reflectivity are typically modeled using the standard Swerling classes. In this paper, we adopt the Swerling I model where the reflectivity is assumed to be  constant within a symbol \cite{xie2020recursive}.}. 
	$\mu_d$ and $\mu_r$ represent the Doppler frequencies corresponding to the direct link and IRS link, respectively, $T_s$ denotes the symbol period, and $\{\mathbf{p}_{r,k}\}_{k=1}^{N_r}$ denote the noise vectors whose elements are drawn independently from a complex Gaussian distribution with zero mean and covariance matrix  $\frac{\sigma_r^2}{N}\mathbf{I}$ \cite{keykhosravi2022ris}.

	\subsection{Geometrical Insights of Doppler Effect}
	\begin{figure*}[!t]
		\centering
		\subfloat[]{\includegraphics[width=2.8in]{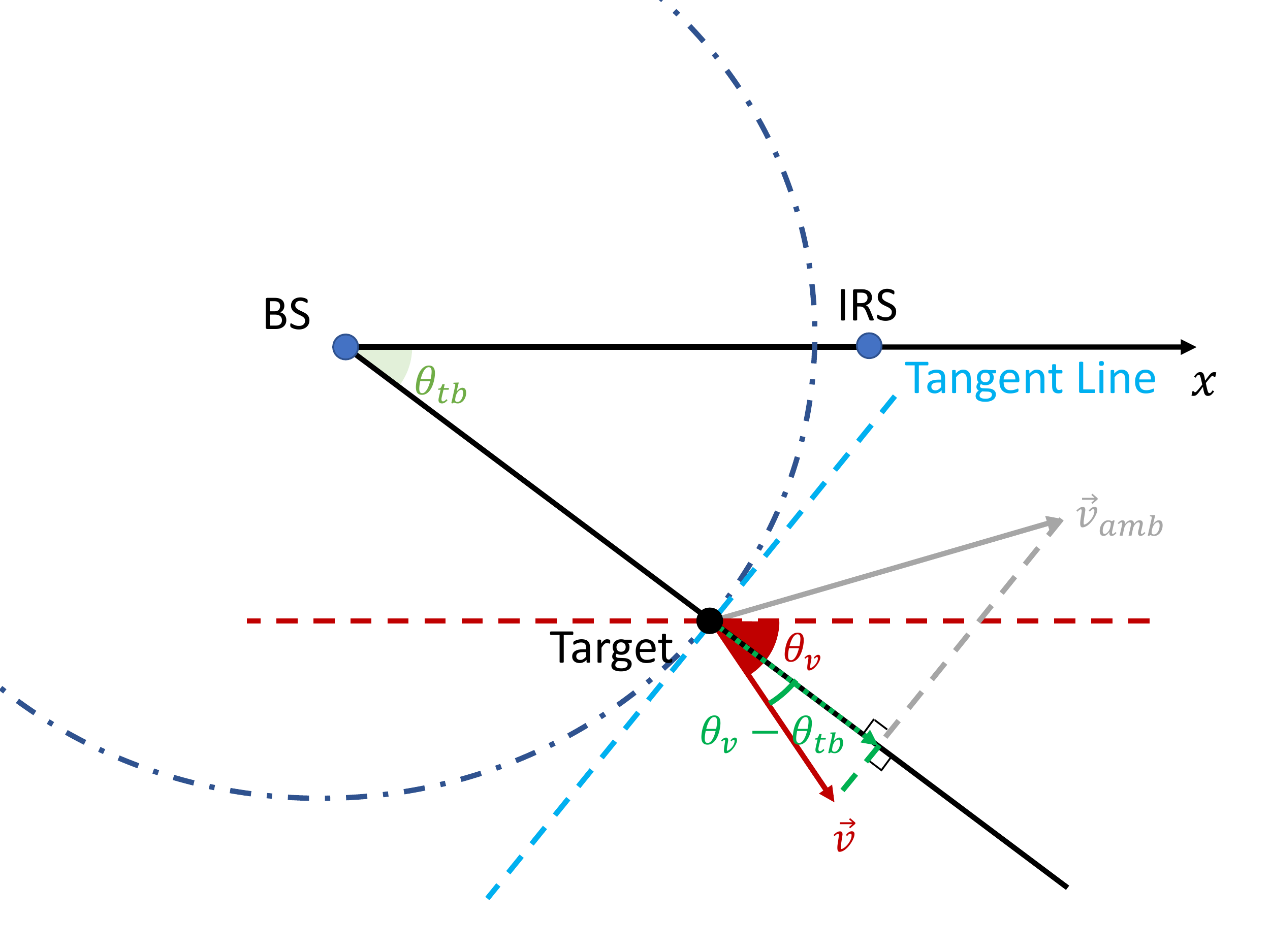}}\
		\subfloat[]{\includegraphics[width=2.8in]{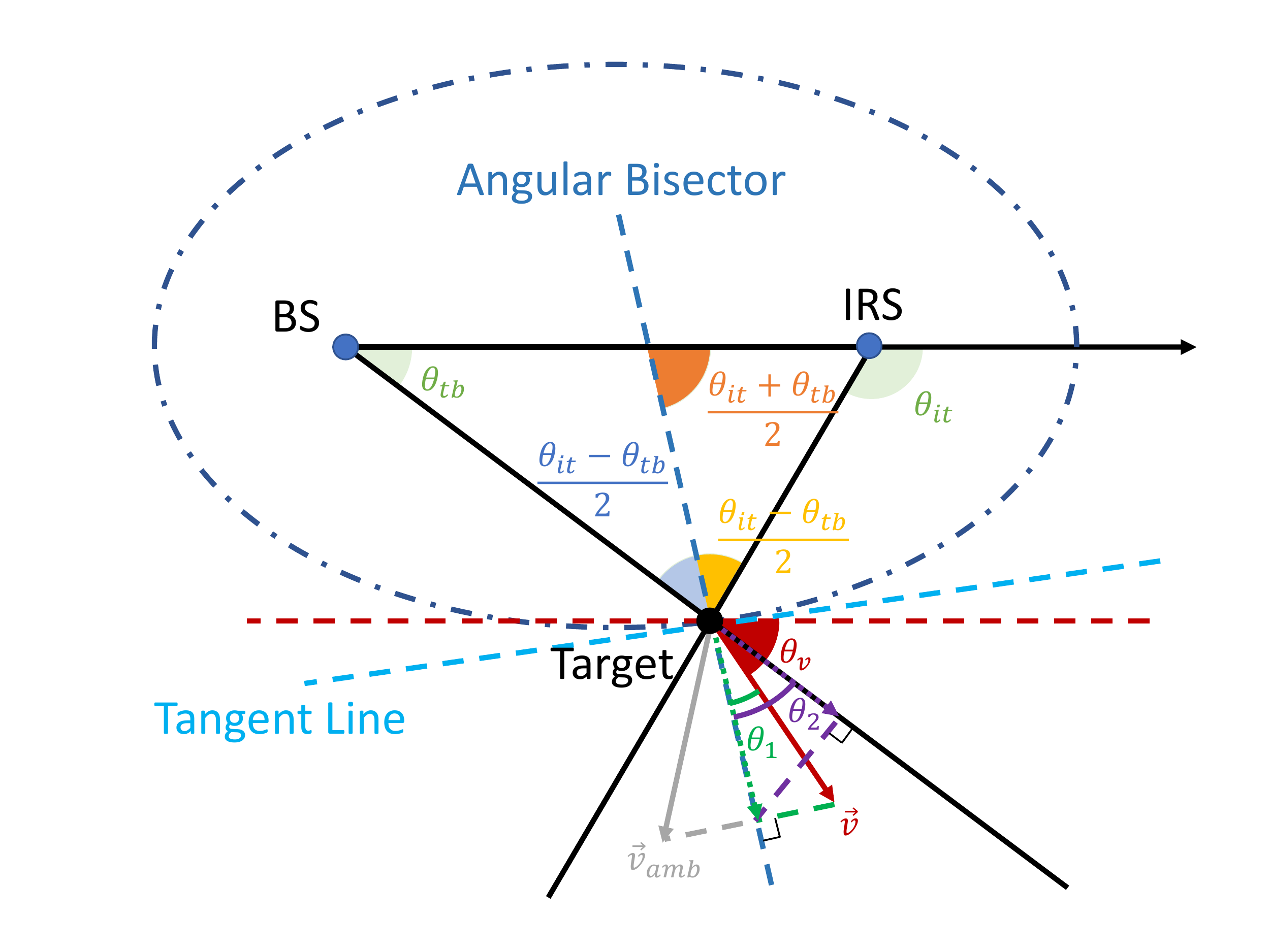}}\\
		\caption{Illustration of Doppler effect. (a)  Direct link; (b) IRS link.}
		\label{fig_illuDop}
	\end{figure*}

	Since the Doppler effect depends on the mobility of the target, we propose to estimate the true velocity $\mathbf{v}$ by exploiting the Doppler frequencies $\{\mu_d,\mu_r\}$.In the following, we first discuss the geometrical insights of the Doppler effect to reveal the relation between true velocity $\mathbf{v}$ and the Doppler frequencies $\{\mu_d,\mu_r\}$.
	
	The Doppler frequency of the relative motion between the BS and the maneuvering target is given by 
	\begin{equation}
		f_{tb}=\frac{v \cos\left(\theta_v -\theta_{tb}\right)}{\lambda},
	\end{equation}
	where $v$ denotes the amplitude of the true velocity, and $\theta_v$ represents the angle between the velocity direction and the $x$-axis, which is marked in red in Fig. \ref{fig_illuDop}. 
	Note that $v \cos\left(\theta_v -\theta_{tb}\right)$ is the projection of velocity on the line across the BS and target, which is marked in dark green in Fig. \ref{fig_illuDop}(a).
	Then, the Doppler frequency corresponding to the direct link is given by
	\begin{equation}\label{mud}
		\mu_{d}=2 f_{tb}=\frac{2 v \cos\left(\theta_v -\theta_{tb}\right)}{\lambda}.
	\end{equation}

	The deployment of the IRS creates another path, which allows the BS to observe the velocity from an additional perspective.
	The Doppler frequency of the relative motion between the IRS and the maneuvering target is given by
	\begin{equation}
		f_{it}=\frac{v \cos\left(\theta_{it}-\theta_v \right)}{\lambda}.
	\end{equation}
	Then, the Doppler frequency corresponding to the IRS link is given by
	\begin{equation}\label{mur}
		\begin{split}
			\mu_{r}&=f_{tb}+f_{it}=\frac{2v}{\lambda}\cos\theta_1\cos\theta_2,
		\end{split}
	\end{equation}
	where
	$\theta_1={\frac{\theta_{tb}+\theta_{it}}{2}-\theta_v}$ and $\theta_2=\frac{\theta_{it} -\theta_{tb}}{2}$ which are marked in dark green and purple in Fig. \ref{fig_illuDop}(b), respectively.

	From (\ref{mur}), $\mu_{r}$ comes from two projections:
	\begin{enumerate}
		\item $\cos\theta_1$: This term projects the velocity to the angular bisector of the angle $\angle_\text{BS-Target-IRS}$. 
		Note that the angular bisector is perpendicular to the tangent line of the ellipse with foci at the BS and IRS.
		\item $\cos\theta_2$: This term projects the projected speed $v\cos \theta_1$ to the line connecting the BS and the target (or the line connecting the IRS and the target). Since the direction of the projected speed $v\cos \theta_1$ is fixed, there will be no ambiguity on the velocity at the projection corresponding to $\cos\theta_2$.
	\end{enumerate}
	
	\textbf{Remark:}
	The conventional mono-static radar can only access $\mu_d$, which corresponds to the radial component of the velocity.
	As a result, the tangent projection will cause ambiguity, e.g., different velocities may yield the same observation at the BS. One example is illustrated in Fig. \ref{fig_illuDop}(a) where two velocity vectors, i.e., $\mathbf{v}_{amb}$ (marked in gray) and $\mathbf{v}$ (marked in red), give the same projection. 
	Similarly, for the IRS link, the velocity ambiguity occurs at the tangent line of the ellipse, as illustrated in Fig. \ref{fig_illuDop}(b). Thus, either of the two links, by itself, will cause velocity ambiguity. Nevertheless, by jointly considering the two links, the true velocity vector can be uniquely determined. Note that we assume that the distance between the BS and the IRS should be comparable to the distance between the target and the BS/IRS.
	
	The following proposition provides the true velocity estimation in the IRS-aided system by jointly leveraging $\mu_d$ and $\mu_r$.
	
	\begin{mypro}\label{prop1}
		The direction and absolute value of the true velocity can be determined uniquely as
		\begin{equation}\label{thetavest}
			\begin{split}
				{\theta}_v=\arctan \left[
				\left(1-\frac{\mu_d}{\mu_r}\right) 
				\cot  \left(\frac{\theta_{it}-\theta_{tb}}{2}\right)
				\right] + \frac{\theta_{tb}+\theta_{it}}{2},
			\end{split}
		\end{equation}
		\begin{equation}\label{vest}
			\begin{split}
				{v}=\frac{\lambda}{2}\cdot \frac{\mu_d}{\cos\left(\theta_v -\theta_{tb}\right)}.
			\end{split}
		\end{equation}
	\end{mypro}
	
	\emph{Proof}: See Appendix. \hfill $\blacksquare$

	\section{True Velocity Estimation}
	According to \emph{Proposition 1}, to estimate the true velocity, we need to estimate $\mu_d$ and $\mu_r$. 
	In this section, we propose a two-stage scheme to estimate the true velocity of the maneuvering target.
	Specifically, in the first stage, we obtain a coarse estimation of $\mu_d$. In the second stage, $\mu_d$ and $\mu_r$ are estimated accurately by utilizing the MODE. The coarse estimation serves two purposes. 
	First, it can be utilized to initialize the iterative algorithm to speed up the iteration in the second stage. Second, the coarse estimation can help match the estimated Doppler frequencies to different links.

	\subsection{Stage I: Coarse Estimation of $\mu_d$}
	We first turn off the IRS to obtain a coarse estimation of $\mu_d$ only based on the direct link. This can be implemented through a receiving beam with zero gain on the direction of IRS. In this case, the system degenerates to a conventional mono-static one. Assume the BS transmits $N_d$ pilot symbols in Stage I. 
    Given the pilot symbol $s_k=1$, the received echo of the $k$-th symbol at the BS is given by
	\begin{equation}\label{ydkdef}
		\begin{split}
			\mathbf{y}_{d,k}=& \alpha_d e^{j2\pi \mu_{d} (k-1) T_{s}} \mathbf{a}(\theta_{tb}) \mathbf{a}^{\mathrm{H}}(\theta_{tb}) \mathbf{f}
			+\mathbf{p}_{d,k},
		\end{split}
	\end{equation}
	where $\{\mathbf{p}_{d,k}\}_{k=1}^{N_d}$ denotes the noise vector whose elements are drawn independently from a complex Gaussian distribution with zero mean and covariance matrix  $\frac{\sigma_d^2}{N}\mathbf{I}$
	
	Define $\mathbf{w}_{d}=\mathbf{a}(\theta_{tb})$ as the combiner corresponding to the received beam which points to the estimated target position. Then the output for the $k$-th symbol is given by 
	\begin{equation}\label{zdkdef}
		\begin{split}
			z_{d,k}= \mathbf{w}_{d}^{\mathrm{H}}\mathbf{y}_{d,k} = \beta_{d} e^{j2\pi \mu_{d} (k-1) T_{s}} & + n_{d,k},
		\end{split}
	\end{equation}
	where $\beta_{d}=\alpha_d \mathbf{w}_d^{\mathrm{H}} \mathbf{a}(\theta_{tb}) \mathbf{a}^{\mathrm{H}}(\theta_{tb}) \mathbf{f}$
	and $n_{d,k} = \mathbf{w}_{d}^{\mathrm{H}}\mathbf{p}_{d,k}$ denotes the combined noise following the Gaussian distribution, i.e., $n_{d,k}\sim \mathcal{CN}(0,\sigma_d^2)$.
	Stacking  $z_{d,k}$ into a vector yields 
	\begin{equation}
		\begin{split}
			\mathbf{z}_d&=\left[ z_{d,1},z_{d,2},\cdots,z_{d,N_d}  \right]^{\mathrm{T}}\in \mathbb{C}^{N_d \times 1}.
			\label{zdvec}
		\end{split}
	\end{equation}
	
	Define the steering vector in the Doppler domain as
	\begin{equation}
		\begin{split}
			\mathbf{a}_f^{(N_d)}(\mu)&\triangleq\left[ 1,e^{-j 2\pi \mu  T_s},\cdots,e^{-j 2\pi \mu (N_d-1) T_s} \right]^{\mathrm{T}}.
			\label{alphavec}
		\end{split}
	\end{equation}
	The coarse estimation of $\mu_d$  is obtained as
	\begin{equation}
		\begin{split}
			\hat{\mu}_{d,\text{coarse}}=\arg \max_{\mu} |\mathbf{z}_d^\mathrm{T}\mathbf{a}_f^{(N_d)}(\mu) |^2.
			\label{coarseestimation}
		\end{split}
	\end{equation}

	\subsection{Stage II: Refined Gridless Estimation of $\mu_d$ and $\mu_r$}
	At this stage, we turn on the IRS to obtain a refined estimation of $\mu_d$ and $\mu_f$. The BS transmits $N_r$ pilot symbols at this stage. 
	Define $\mathbf{w}_{r}=\mathbf{a}(\theta_{tb})+\mathbf{a}({\theta}_{bi})$ as the combiner illuminated at the estimated target position and IRS, simultaneously, where $\theta_{bi}$ denotes the direction of the IRS with respect to the BS. 
	Then, the output for the $k$-th symbol is given by
	\begin{equation}\label{zrkdef}
		\begin{split}
			z_{r,k}= \mathbf{w}_{r}^{\mathrm{H}}\mathbf{y}_{r,k} = \beta_{r} e^{j2\pi \mu_{r} (k-1) T_{s}} &+ n_{r,k},
		\end{split}
	\end{equation}
	where  $\beta_{r}=\alpha_r \mathbf{w}_r^{\mathrm{H}} \mathbf{G} \mathbf{\Psi} \mathbf{b}(\theta_{it}) \mathbf{a}^{\mathrm{H}}(\theta_{tb}) \mathbf{f}$
	and $n_{r,k} = \mathbf{w}_{r}^{\mathrm{H}}\mathbf{p}_{r,k}$ denotes the combined noise, which follows the Gaussian distributions, i.e., $n_{r,k}\sim \mathcal{CN}(0,\sigma_r^2)$.
	
	Before proceeding, we first construct a set of  $P \times 1$ signal vectors by rearranging $\{z_{r,k}\}_{k=1}^{N_r}$. By doing that, the Doppler frequency estimation problem can be solved by the MODE.
	Given a positive integer $P$, the signal vector is defined as
	\begin{equation} 
		\begin{split}
			\mathbf{z}_{r,k}=\left[ z_{r,k},z_{r,k-1},\cdots,z_{r,k-P+1} \right]^{\textrm{T}} &\in \mathbb{C}^{P\times 1}.
		\end{split}
		\label{bfzrk}
	\end{equation}
	{From (\ref{zrkdef}),} (\ref{bfzrk}) can be rewritten as
	\begin{equation} 
		\mathbf{z}_{r,k}=\mathbf{A}\mathbf{x}_{r,k}+\mathbf{n}_{r,k},
		\label{yy1kmat}
	\end{equation}
	where
	\begin{equation}
		\mathbf{A}=\left[ \mathbf{a}_f^{(P)}(\mu_d),\mathbf{a}_f^{(P)} (\mu_r) \right],
		\label{Amat}
	\end{equation}
	\begin{equation}
		\begin{split}
			\mathbf{x}_{r,k}=[\beta_{d}e^{j 2\pi \mu_d (k-1)  T_s},\beta_{r}  e^{j 2\pi \mu_r (k-1)  T_s}  ]^{\textrm{T}},
			\label{svec}
		\end{split}
	\end{equation}
	\begin{equation}
		\mathbf{n}_{r,k}=\left[ n_{r,k},n_{r,k-1},\cdots,n_{r,k-P+1} \right]^{\textrm{T}}.
		\label{v1}
	\end{equation}
	The covariance matrix of $\mathbf{z}_{r,k}$ is
	\begin{equation}
		\begin{split}
			\mathbf{R}_r=\mathbb{E} \left\lbrace \mathbf{z}_{r,k} \mathbf{z}_{r,k}^{\textrm{H}} \right\rbrace
			=\mathbf{A}\mathbf{\Phi}\mathbf{A}^{\textrm{H}}+\sigma_r^2 \mathbf{I},
			\label{Rcov}
		\end{split}
	\end{equation}
	where $\mathbb{E}{\{\cdot\}}$ denotes the expectation over $k$ and $\mathbf{\Phi}=\mathbb{E}{\{\mathbf{x}_{r,k}\mathbf{x}_{r,k}^\mathrm{H}\}}$. 
	The eigenvalue decomposition of $\mathbf{R}_r$ gives 
	\begin{equation}
		\begin{split}
			\mathbf{R}_r=\sum_{i=1}^P \lambda_i \mathbf{g}_i \mathbf{g}_i^\mathrm{H},
			\label{eigendis}
		\end{split}
	\end{equation}
		where $\lambda_1 \geq \lambda_2 \geq \lambda_3 = \cdots = \lambda_P=\sigma_r^2$ are the eigenvalues, {$\mathbf{G}_s=\left[\mathbf{g}_1,\mathbf{g}_2\right]\in \mathbb{C}^{P\times 2}$ and  $\mathbf{G}_N=\left[\mathbf{g}_{3},\mathbf{g}_{4},\cdots,\mathbf{g}_{P}\right]\in \mathbb{C}^{P \times (P-2)}$ correspond to the signal and noise subspace, respectively.}   
		Note that the linear span of $\mathbf G_s$ is the same as
		that of $\mathbf{A}$, i.e.,
		\begin{equation}
			\text{span}(\mathbf{G}_s)=\text{span}(\mathbf{A}).
			\label{spanGs}
		\end{equation}

		Note that the maximum likelihood estimate (MLE) of $\bm{\mu}=\{\mu_d,\mu_r\}$ depends on $\mathbf{R}_r$, which is unknown in practice. 
		To address this, we first obtain the estimated eigenvectors $\{\widehat{\mathbf{g}}_i\}_{i=1}^P$ and the estimated eigenvalues $\{\widehat{\lambda}_i\}_{i=1}^P$ by performing the eigenvalue decomposition on the sample covariance matrix of $\{\mathbf{z}_{r,k}\}_{k=P}^{N_r}$, i.e.,  
		\begin{equation}
			\begin{split}
				\widehat{\mathbf{R}}_{r,\text{SCM}}\triangleq\frac{1}{N_r-P+1} \sum_{k=P}^{N_r}\mathbf{z}_{r,k} \mathbf{z}_{r,k}^{\textrm{H}}.	
			\end{split}
		\end{equation}
		Then, an approximated maximum likelihood estimate (MLE) of $\bm{\mu}=\{\mu_d,\mu_r\}$ is given by \cite{57542}
		\begin{equation}
			\begin{split}
				{\bm{\widehat{\mu}} =\arg\min_{\bm{\mu}} \mathrm{tr} \left\{ \left(\mathbf{I}-\Phi_\mathbf{A}(\bm{\mu})\right) \mathbf{\widehat{G}}_s \widehat{\mathbf \Gamma} \mathbf{\widehat{G}}_s^{\rm{H}}\right\},}
				\label{mlest}
			\end{split}
		\end{equation}
		where $\mathrm{tr}(\cdot)$ denotes the trace, $\Phi_\mathbf{A}(\bm{\mu})=\mathbf{A} \left( {\mathbf{A}}^{\mathrm{H}} \mathbf{A} \right)^{-1} {\mathbf{A}}^{\mathrm{H}}$
		represents the orthogonal projector onto the range space of  $\mathbf{A}$, and
		$\mathbf{\widehat{G}}_s =[\widehat{\mathbf{g}}_1,\widehat{\mathbf{g}}_2]$, while
		\begin{equation}
			\begin{split}
				\widehat{\mathbf \Gamma} \triangleq {\rm{diag}} \left\{ \widehat{\lambda}_1^{-1}\left(\widehat{\lambda}_1-\widehat{\sigma}_r^2\right)^2,
				\widehat{\lambda}_2^{-1}\left(\widehat{\lambda}_2-\widehat{\sigma}_r^2\right)^2\right\}, 
				\label{gammahat}
			\end{split}
		\end{equation}
		{and $\widehat{\sigma}_r^2=\sum_{i=3}^{P} \widehat{\lambda}_i/(P-2)$}. 
		
		However, the problem in (\ref{mlest}) does not have an analytical solution and is thus hard to solve directly. 	
		To solve (\ref{mlest}), we then consider introducing some tractable intermediate variables, based on which $\mu_d$ and $\mu_r$ can be estimated.
		Construct a polynomial
		\begin{equation}
			\begin{split}
				\left( 1-e^{j 2\pi \mu_d T_{s}}\omega \right)\left( 1-e^{j 2\pi \mu_r T_{s}}\omega \right) =1+c_1 \omega+c_2 \omega^2, 
				\label{poly}
			\end{split}
		\end{equation}
		which has $2$ roots in the complex plane. 
		It can be validated that $c_1=-\left(e^{j 2\pi \mu_d T_{s}}+e^{j 2\pi \mu_r T_{s}}\right)$ and $c_2=e^{j 2\pi (\mu_d+\mu_r) T_{s}}$.
		Clearly, $\omega_d=e^{-j 2\pi \mu_d T_{s}}$ and $\omega_r=e^{-j 2\pi \mu_r T_{s}}$ are the roots of this polynomial. Then, the strategy of solving (\ref{mlest}) is to obtain the intermediate variables $c_1$ and $c_2$, based on which $\mu_d$ and $\mu_r$ can be estimated, instead of estimating $\mu_d$ and $\mu_r$ directly.

		Define a matrix
		\begin{equation}
			\begin{split}
				\mathbf{C}=\left[
				\begin{matrix}
					1 & c_1 &  c_2 & 0 & \cdots & 0\\
					0 & 1 & c_1 & c_2 & \cdots & 0\\
					\vdots & \vdots &  \vdots & \vdots & \ddots & 0 \\
					0& \cdots & 0 & 1 & c_1 &  c_2
				\end{matrix}\right]\in \mathbb{C}^{(P-2)\times P}.
				\label{Bbb}
			\end{split}
		\end{equation}
		From the definition of $\mathbf A$ in (\ref{Amat}), we have  $\mathbf{C}\mathbf{A}= \mathbf 0$,
		which indicates
		\begin{equation}
			\begin{split}
				\mathbf{I}-\Phi_\mathbf{A}(\bm{\mu})=\Phi_\mathbf{C^{\mathrm{H}}}=\mathbf{C}^{\mathrm{H}} \left( \mathbf{C} \mathbf{C}^{\mathrm{H}} \right)^{-1} \mathbf{C}.
				\label{orth}
			\end{split}
		\end{equation}

		Substituting (\ref{orth}) into (\ref{mlest}) produces
		\begin{equation}
			\begin{split}
				&\mathrm{tr} \left\{ \mathbf{C}^{\mathrm{H}} \left( \mathbf{C} \mathbf{C}^{\mathrm{H}} \right)^{-1} \mathbf{C} \mathbf{\widehat{G}}_s \widehat{\mathbf \Gamma} \mathbf{\widehat{G}}_s^{\rm{H}} \right\}\\
				= &\mathrm{tr} \left\{  \left( \mathbf{C} \mathbf{C}^{\mathrm{H}} \right)^{-1} \mathbf{C} \mathbf{\widehat{G}}_s \widehat{\mathbf \Gamma} \mathbf{\widehat{G}}_s^{\rm{H}}\mathbf{C}^{\mathrm{H}} \right\}\\
				=&\mathrm{vec} \left\{\mathbf{C} \mathbf{\widehat{G}}_s\right\}^{\mathrm{H}}
				\mathbf{W}
				\mathrm{vec}\left\{\mathbf{C} \mathbf{\widehat{G}}_s\right\}.
				\label{mlest2}
			\end{split}
		\end{equation}
		where $\mathrm{vec}(\cdot)$ denotes the vectorization operator and 
		\begin{equation}\label{W}
			\begin{split}
				\mathbf{W}= \widehat{\mathbf \Gamma} \otimes \left( \mathbf{C} \mathbf{C}^{\mathrm{H}} \right)^{-1}.
			\end{split}
		\end{equation}
		
		Recall from (\ref{poly}) that $\mathbf{c} \triangleq \left[ c_1,c_2 \right]^{\mathrm{T}}$ is related to $\bm{\mu} $. 
		Then, the MLE of $\mathbf{c}$ is given by
		\begin{equation}
			\begin{split}
				\mathbf{\widehat{c}}=\arg\min_{\mathbf{c}}
				\mathrm{vec} \left\{\mathbf{C} \mathbf{\widehat{G}}_s\right\}^{\mathrm{H}}
				\mathbf{W}
				\mathrm{vec}\left\{\mathbf{C} \mathbf{\widehat{G}}_s\right\}.
				\label{mlest3}
			\end{split}
		\end{equation}
		
		Define matrix
		\begin{equation}
			\begin{split}
				\mathbf{\widehat{\Psi}}_j \triangleq \left[
				\begin{matrix}
					\widehat{g}_{2,j} & \widehat{g}_{3,j} \\
					\widehat{g}_{3,j} & \widehat{g}_{4,j} \\
					\vdots & \vdots \\
					\widehat{g}_{P-1,j} & \widehat{g}_{P,j} \\
				\end{matrix}\right]\in \mathbb{C}^{(P-2)\times 2},\quad j=1,2,
				\label{psil}
			\end{split}
		\end{equation}
		and vector
		\begin{equation}
			\begin{split}
				\mathbf{\widehat{q}}_j \triangleq -\left[ \widehat{g}_{1,j},\widehat{g}_{2,j},\cdots,\widehat{g}_{P-2,j} \right]^{\mathrm{T}},\quad j=1,2,
				\label{etal}
			\end{split}
		\end{equation}
		where 
		$\widehat{g}_{i,j}$ denotes the $(i,j)$-th entry of $\widehat{ \mathbf{G} }_s$.
		We have
		\begin{equation}
			\begin{split}
				\mathrm{vec} \left\{\mathbf{C} \mathbf{\widehat{G}}_s\right\}
				&=\mathbf{\widehat{\Psi}} \mathbf{c}-\mathbf{\widehat{q}},
				\label{vecbgs}
			\end{split}
		\end{equation}
		where
		\begin{equation}\label{psihat}
			\mathbf{\widehat{\Psi}}=\left[ \mathbf{\widehat{\Psi}}_1^{\mathrm{T}},\mathbf{\widehat{\Psi}}_2^{\mathrm{T}} \right]^{\mathrm{T}},
		\end{equation}	
		\begin{equation}\label{qhat}
			\mathbf{\widehat{q}}=\left[ \mathbf{\widehat{q}}_1^{\mathrm{T}},\mathbf{\widehat{q}}_2^{\mathrm{T}} \right]^{\mathrm{T}}.
		\end{equation}
		Substituting (\ref{vecbgs}) into (\ref{mlest3}) leads to
		\begin{equation}
			\begin{split}
				\mathbf{\widehat{c}}&=\arg\min_{\mathbf{c}}
				\left(\mathbf{\widehat{\Psi}} \mathbf{c}-\mathbf{\widehat{q}} \right)^{\mathrm{H}}
				\mathbf{W}
				\left(\mathbf{\widehat{\Psi}} \mathbf{c}-\mathbf{\widehat{q}} \right).
				\label{mlestWLS}
			\end{split}
		\end{equation}
		Note that (\ref{mlestWLS}) is a weighted least squares problem whose solution is given by     
		\begin{equation}
			\begin{split}
				\mathbf{\widehat{c}}_\dagger=\left(
				\mathbf{\widehat{\Psi}}^{\mathrm{H}}
				\mathbf{W}
				\mathbf{\widehat{\Psi}} \right)^{-1}
				\mathbf{\widehat{\Psi}}^{\mathrm{H}}
				\mathbf{W}
				\mathbf{\widehat{q}}.
				\label{WLSsolution}
			\end{split}
		\end{equation}
		However, $\bf{W}$ also depends on 
		$\mathbf{C}$ through (\ref{W}) and (\ref{Bbb}).
		In order to tackle this challenge, we propose an iterative algorithm to find the solution, which is summarized in \textbf{Algorithm \ref{algxx}}.	
		\begin{algorithm}[t]
			\caption{MODE-based velocity estimation algorithm} 
			\label{algxx} 
			Initialize $\widehat{\mathbf \Gamma}$,$\mathbf{\widehat{\Psi}}$, and $\mathbf{\widehat{q}}$ via (\ref{gammahat}), (\ref{psihat}), and (\ref{qhat}).\\
			Initialize $\mathbf{\widehat{c}}_{(0)}=[-\left(e^{j 2\pi \hat{\mu}_{d,\text{coarse}} T_{s}}+1\right),e^{j 2\pi \hat{\mu}_{d,\text{coarse}} T_{s}}]^\mathrm{T}$, $t=0$.\\
			Repeat: 
			\begin{enumerate}
				\item {U}pdate $\widehat{\mathbf{C}}_{(t+1)}$ using $\mathbf{\widehat{c}}_{(t)}$, via (\ref{Bbb})
				\item {U}pdate $\mathbf{\widehat{c}}_{(t+1)}$ using $\widehat{\mathbf{C}}_{(t+1)}$, via (\ref{WLSsolution})
				\item $t\gets t+1$.
			\end{enumerate}
			Until some termination condition is met.
		\end{algorithm}

		Once $\mathbf{\widehat{c}}$ is obtained, we can solve the equation 
		\begin{equation}
			\begin{split}
				1+\widehat{c}_1 \omega+\widehat{c}_2 \omega^2=0
				\label{polqua}
			\end{split}
		\end{equation}
		to obtain two roots
		\begin{equation}
			\omega_{1}=\frac{-\widehat{c}_1 -\sqrt{\widehat{c}_1^2-4\widehat{c}_2}}{2\widehat{c}_2}, 	\omega_{2}=\frac{-\widehat{c}_1 +\sqrt{\widehat{c}_1^2-4\widehat{c}_2}}{2\widehat{c}_2}.
		\end{equation}
		The corresponding $\mu_{1}$ and $\mu_{2}$ are obtained as
		\begin{equation}
			\hat{\mu}_{1}=\frac{1}{2\pi T_s}\mathcal{P} (\omega_{1}), 	\hat{\mu}_{2}=\frac{1}{2\pi T_s}\mathcal{P}(\omega_{2}),
		\end{equation}
		where $\mathcal{P}(\cdot)$ denotes the phase of a complex number.

		We can obtain the refined estimates of $\mu_d$ and $\mu_r$, i.e.,
		\begin{equation}
			\begin{split}
				&\hat{\mu}_d=\mathop{\arg\min}_{\mu\in \{\hat{\mu}_1,\hat{\mu}_2\}}|\hat{\mu}_{d,\text{coarse}}-\mu|,\\
				&\hat{\mu}_r=\mathop{\arg\max}_{\mu\in\{\hat{\mu}_1,\hat{\mu}_2\}}|\hat{\mu}_{d,\text{coarse}}-\mu|.
			\end{split}
		\end{equation}
		By exploiting \emph{Proposition 1}, the direction and absolute value of true velocity can be estimated.

		\section{Simulation Results}
		In this section, we show the performance of the proposed velocity estimator. The number of antennas at the BS is set as $N=16$ and the number of elements at the IRS is set as $M=32$. The carrier frequency is $f_c=3$ GHz and the symbol period is $T_{s}=0.5$ ms. The number of symbols at stages I and II are $N_d =16$ and $N_r=16$, respectively.
		Assume that the BS and IRS are located at $(0,0)$ and $(20,0)$, respectively, where all coordinates are in meters. The direction of the target with respect to the BS and IRS are $\theta_{tb}=30^\circ$ and $\theta_{it}=120^\circ$, respectively. 
		$\mathbf{G}$ is modeled as a Rician channel comprising a LOS path and a number of NLOS paths. The number of paths for $\mathbf{G}$ is set as $N_p=3$, where the AoA and AoD parameters are uniformly generated from $[-\pi/2,\pi/2]$. 
		The Rician factor is set as $13.2$ dB.
		The signal-to-noise ratio (SNR) of the direct link is defined as $\text{SNR}=\frac{\mathbb{E}(\alpha_d^2)}{\sigma_r^2}$.

		\begin{figure}[!t]
			\centering
			\includegraphics[width=3.0in]{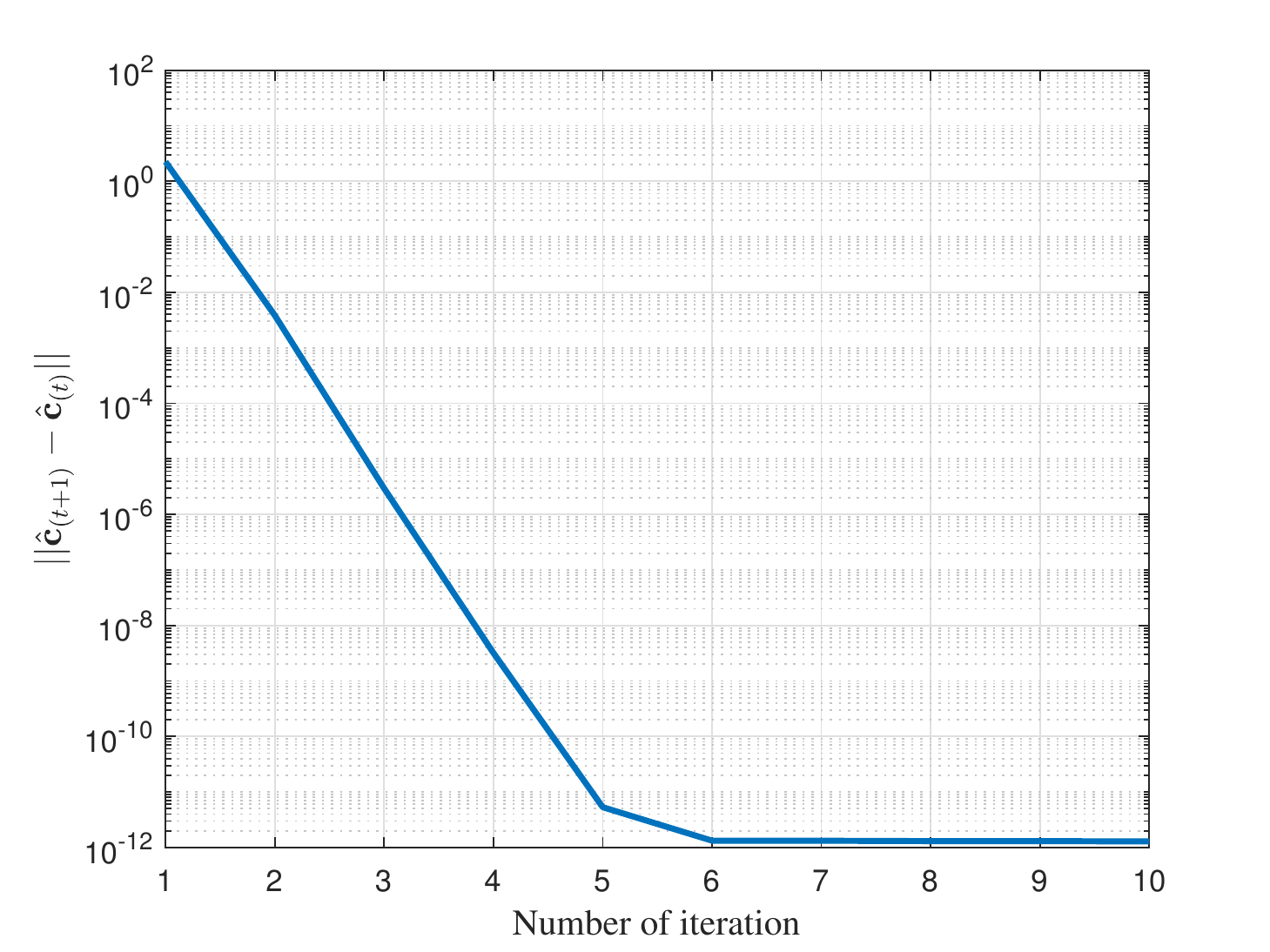}
			\caption{Convergence of the proposed MODE-based method. }
			\label{fig_iter}
		\end{figure}

		\begin{figure}[!t]
			\centering
			\includegraphics[width=3.0in]{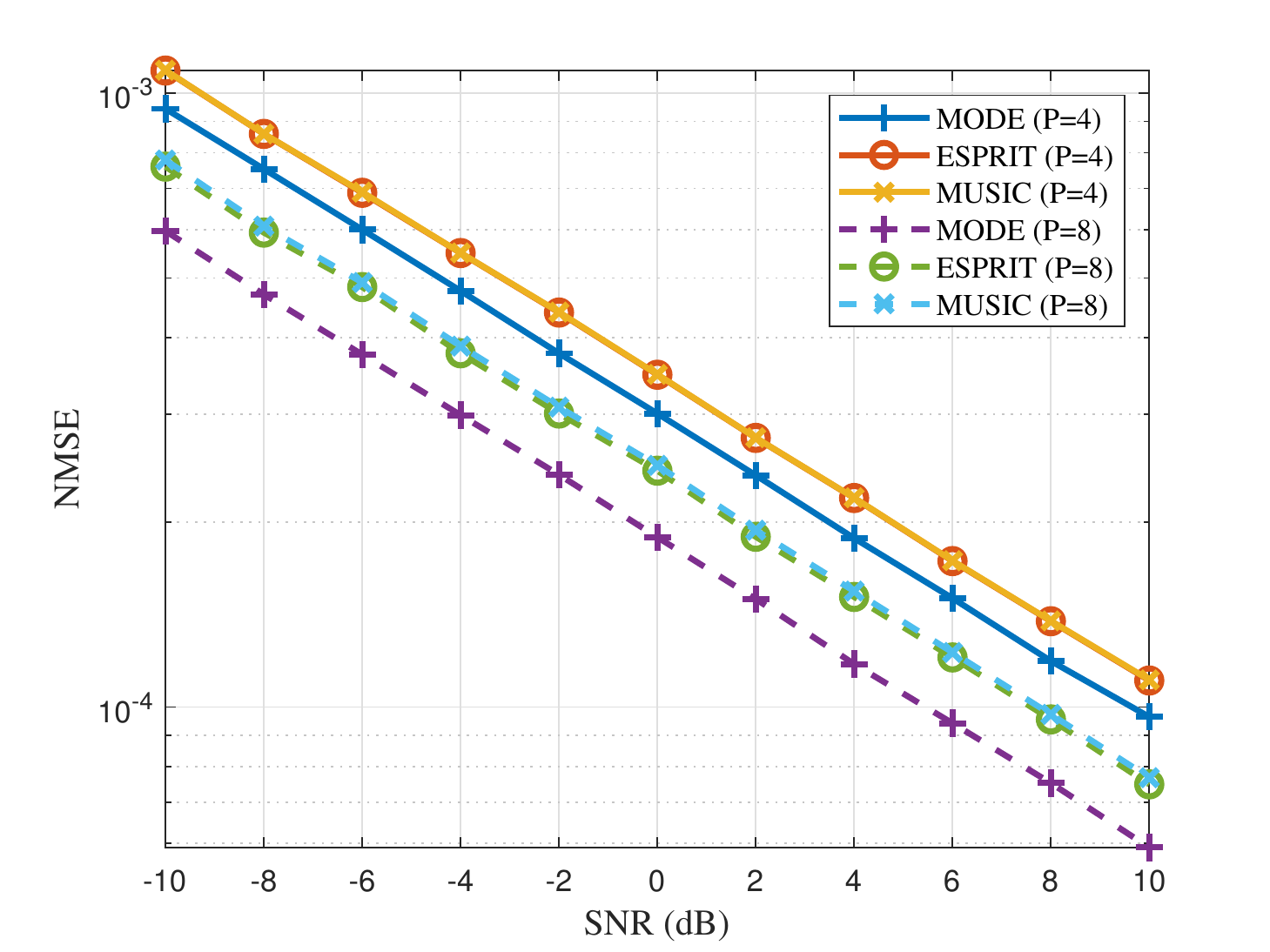}
			\caption{NMSE performance versus SNR. }
			\label{fig_MSE}
		\end{figure}
		
		We first show the convergence performance of the proposed MODE-based method, which involves an iterative process.
		For that purpose, we choose the average discrepancy of $\widehat{\mathbf{c}}_{(t)}$ between two adjacent iterations as the metric, i.e.,
		$\mathcal{D}_{(t)}=||\widehat{\mathbf{c}}_{(t+1)}-\widehat{\mathbf{c}}_{(t)}||.$ For each abscissa, $10,000$ Monte-Carlo trials are carried out.
		From Fig. \ref{fig_iter}, we can see that the iteration  converges in about ten rounds.
		
		We compare the proposed MODE-based method with two gridless alternatives, i.e., root-MUSIC \cite{1143830} and root-ESPRIT \cite{32276}.
		To quantify the estimation accuracy, we evaluate the normalized mean squared error (NMSE), i.e., $	\text{NMSE}=\sqrt{\frac{1}{N_{mc}}{\sum_{i=1}^{N_{mc}}\frac{||\mathbf{v}-\widehat{\mathbf{v}}_i||^2}{||\mathbf{v}||^2}}},
		$
		where $N_{mc}$ denotes the number of Monte-Carlo trials and $\widehat{\mathbf{v}}_i$ denotes the estimation result in the $i$-th Monte-Carlo trial.
		
		The NMSEs of different methods under different SNRs are shown in Fig. \ref{fig_MSE}.
		The velocity of target is $40$ m/s with $\theta_v = 60^\circ$ and $N_{mc}=10,000$.
		From Fig. \ref{fig_MSE}, we observe that the true velocity can be recovered with a considerable accuracy.
		As SNR increases, the NMSEs of all these methods decrease.		
		Furthermore, the proposed MODE can achieve higher accuracy than MUSIC and ESPRIT. 
		This is because the MODE can achieve the minimum estimation variance among the class of methods based on the eigenspace of the sample covariance matrix including MUSIC and ESPRIT \cite{57542}. 
		\begin{figure}[!t]
			\centering
			\includegraphics[width=3.0in]{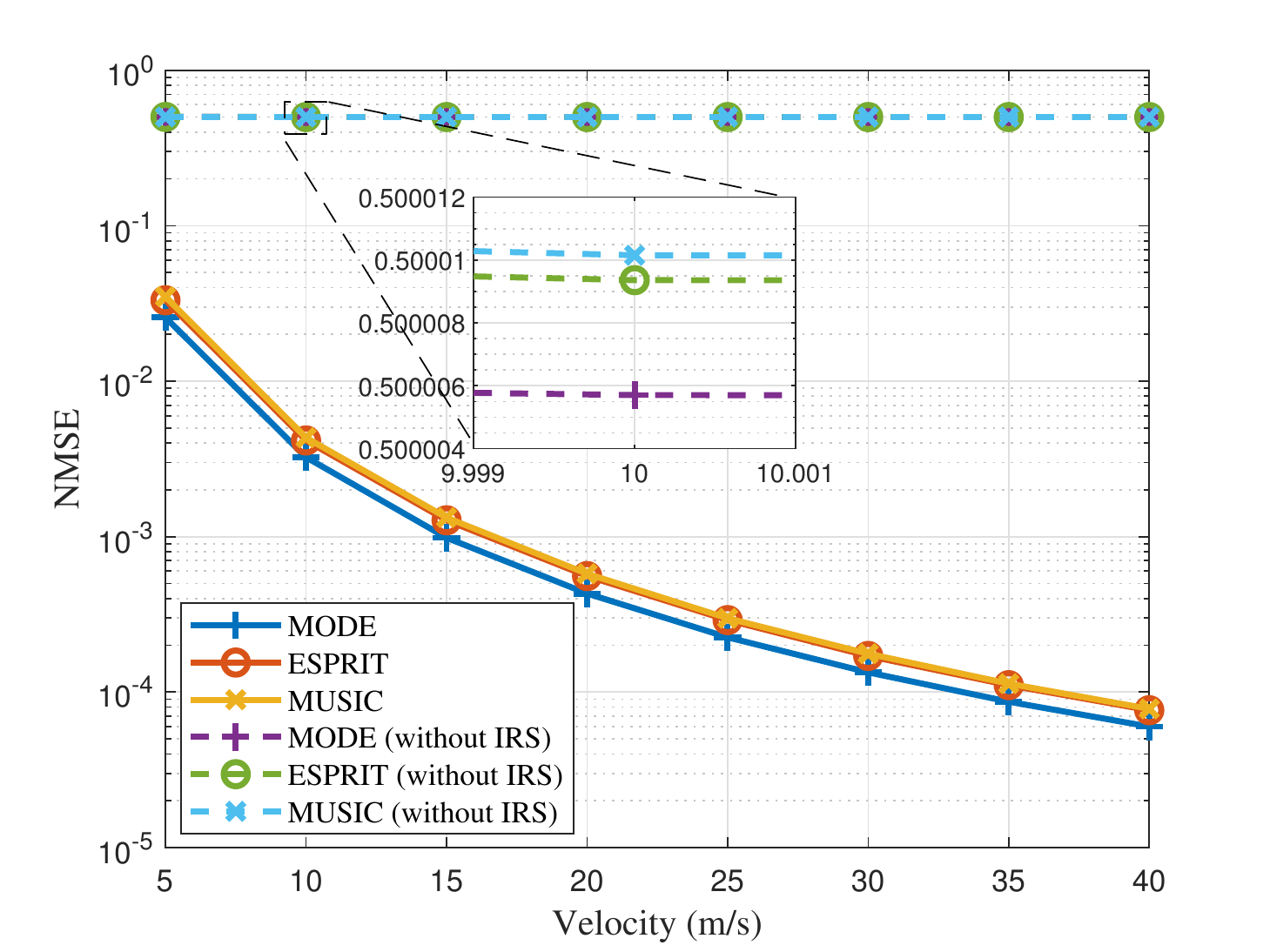}
			\caption{NMSE performance versus velocity $|\mathbf{v}|$. }
			\label{fig_v}
		\end{figure}
		
		Next, we show the NMSE performance under different target velocities in Fig. \ref{fig_v}. 
		We set $\theta_v=60^\circ$, $N_{mc}=10,000$, and $\text{SNR}=0$ dB.  From Fig. \ref{fig_v}, we can see that the NMSE performance is significantly improved with the deployment of the IRS. 
		This is because only a radial projection of the true velocity can be estimated for a single BS without the IRS, such that the tangent projection of the true velocity dominates the estimation error. 
		Meanwhile, with the IRS, the NMSE performance improves as the target velocity increases.
		The reason for this phenomenon requires a further theoretical study on the Cram\'{e}r-Rao bound, which will be left to our future work. 
		Moreover, the gap between the curves with and without the IRS becomes larger as $v$ increases, which indicates that the benefit of the deployment of IRSs will be more pronounced in the high-mobility case.
		
		\section{Conclusion}
		This paper considered the true velocity estimation for a maneuvering target with the assistance of an IRS. Thanks to the extra link provided by the IRS, the BS can observe the true velocity from two different perspectives.
		In particular, we proposed a two-stage algorithm to estimate the Doppler frequency of the direct link and IRS link, based on which the true velocity recovered.
		Compared with existing approaches, the proposed approach achieved much better estimation performance. The work in this paper represents a small step to fully exploit the power of IRS in ISAC applications.

		\appendix
		
		\section*{Proof of Proposition 1}
		(\ref{vest}) can be directly obtained from (\ref{mud}), thus we only prove (\ref{thetavest}) in the following.
		Recalling the definition of $\theta_1$ and $\theta_2$, we have 
		\begin{equation}
			\theta_v-\theta_{tb}=\theta_2-\theta_1.
		\end{equation}
		Then, (\ref{mud}) can be reformulated as
		\begin{equation}
			\mu_d=\frac{2v\cos\left(\theta_2-\theta_1\right)}{\lambda}.
		\end{equation}
		
		According to (\ref{mur}), we have
		\begin{equation}\label{muddevidemur}
			\begin{split}
				\frac{\mu_d}{\mu_r}&=\frac{\cos\left(\theta_2-\theta_1\right)}{\cos\theta_1\cos\theta_2}=1+\tan \theta_1 \tan \theta_2.
			\end{split}
		\end{equation}
		
		By rearranging (\ref{muddevidemur}), we have
		\begin{equation}
			\theta_v-\frac{\theta_{tb}+\theta_{it}}{2}=-\theta_1 = \arctan \left[\left(1-\frac{\mu_d}{\mu_r}\right)\cot \theta_2\right],
		\end{equation}
		which completes the proof of (\ref{thetavest}).   \hfill $\blacksquare$

	\end{document}